\documentclass[twocolumn, pra] {revtex4}
\usepackage[dvips]{graphics,color,floatflt,epsfig}
\DeclareGraphicsExtensions{.eps,.png}
\usepackage[letterpaper,colorlinks,breaklinks,bookmarks]{hyperref}
\begin{document}
\renewcommand{\Re}{\hbox{Re\,}}
\renewcommand{\Im}{\hbox{Im\,}}
\newcommand{\manual}{rm}        
\def\dfrac{\displaystyle\frac}
%
%
\sloppy \thispagestyle{empty}
\title{Coherence-controlled transparency and far-from-degenerate parametric gain in a
strongly-absorbing Doppler-broadened medium}
\author{Thomas F. George}
\affiliation{Office of the Chancellor / Departments of Chemistry and Physics \&
Astronomy\\
University of Wisconsin-Stevens Point, Stevens Point, WI 54481-3897, USA}
\author{A. K. Popov}
\affiliation{Institute for Physics, Russian Academy of Sciences, Krasnoyarsk, 660036,
Russia}
\author{Received by OPTICS LETTERS March 24, 2000}
\affiliation{}
\begin{abstract}An inversionless amplification  of  anti-Stokes radiation
above the oscillation threshold in an optically-dense far-from-degenerate
double-$\Lambda$ Doppler-broadened medium accompanied by Stokes gain is predicted.
The outcomes are  illustrated with numerical simulations applied to sodium dimer
vapor. Optical switching from absorption to gain via transparency controlled by a
small variation of the medium and of the driving radiation parameters which are at a
level less than one photon per molecule is shown.  Related video/audio clips see in:
A.K. Popov, S.A. Myslivets, and T.F. George, Optics Express {\bf 7}, No 3, 148
(2000), \href{http://epubs.osa.org/oearchive/source/22947.htm}
{http://epubs.osa.org/oearchive/source/22947.htm}, or download:
\href{http://kirensky.krasn.ru/popov/opa/opa.htm}{http://kirensky.krasn.ru/popov/opa/opa.htm}

{\it OCIS codes:} 020.1670; 030.1670; 190.0190; 190.4970
\end{abstract}
\maketitle The concept of quantum coherence and interference in multilevel schemes as
an origin of difference in probabilities of absorption and induced emission
\cite{Rau} plays an important role in optical physics. A mechanism for achieving
amplification without bare-state population inversion (AWI) in an optical transition
based on this concept was explicitly proposed, numerically illustrated for the
$V$-type scheme of Ne transitions \cite{AWI}, and experimentally  proved \cite{Bet}.
Coherence and interference effects have been extensively explored to manipulate
energy level populations, nonlinear-optical response, refraction, absorption and
amplification of optical radiation in resonant media over the past decade (for a
review, see Ref. 4). Recent theoretical and experimental investigations have
demonstrated a slowing down of the light group speed to a few meters per second,
highly efficient frequency conversion, optical switches, and potential sources of
squeezed quantum state light for quantum information processing on this basis
\cite{Har,QN}.  In this Letter we propose a scheme for coherent quantum control (CQC)
that enables one to form upconverted AWI without incoherent excitation in an
initially strongly absorbing optically dense medium as well as to accomplish optical
switching. All this is shown to be possible by application of driving fields at the
level of one photon per several molecules. Unlike other major approaches, the proposed
scheme  does not require coherent population trapping. In contrast to the research
reported in Ref. 7, for which the frequencies of all the coupled fields are almost
equal, we consider far-from-degenerate CQC, for which the features of an optically
dense Doppler-broadened medium play a crucial role. The interplay of absorption,
gain, four-wave mixing (FWM) and their interference at inhomogeneously broadened
transitions is shown to result in substantial enhancement of the gain by proper
resonance detuning, which was not realized in a recent experiment \cite{Hin}.

Let us consider four-level scheme [Fig. 1(a)]. The fields $E_1(\omega_{1})$ and
$E_3(\omega_{3})$  are driving, whereas $E_4(\omega_{4})$ and $E_2(\omega_{2})$ are
weak probes. The problem under consideration reduces to the solution of a set of the
coupled equations  for four waves, $(E_i/2)\, \exp [i(k_iz-\omega_i t)] + c.c.\, \,
(i= 1...4)$, copropagating in an optically thick medium:
\begin{eqnarray}
&{dE_{4,2}(z)}/{dz}= i\sigma_{4,2} E_{4,2}+i\tilde\sigma_{4,2}
E_1E_3E_{2,4}^*, &\label{e12}\\
&{dE_{1,3}(z)}/{dz} = i\sigma_{1,3} E_{1,3}+
i\tilde\sigma_{1,3}E_4E_2E_{3,1}^*. &\label{e13}
\end{eqnarray}
Here $\omega_4+\omega_2 =\omega_1+\omega_3$, $k_j$  are wave numbers in vacuum;
$\tilde\sigma_j=-2\pi k_j \tilde\chi_j $, and $\sigma_j=-2\pi k_j \chi_j = \delta
k_j+i{\alpha_j}/{2}$; and $\tilde\chi_j$, $\alpha_j$, and $\delta k_j$ are
intensity-dependent cross-coupling susceptibilities, absorption indices, and
dispersion parts of $k_j$.
\begin{figure}[htb]
\epsfysize=4.5cm \center{\leavevmode\epsfbox{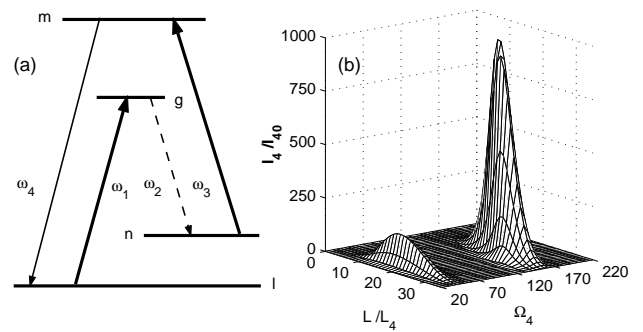}} \caption{(a) Energy levels and
(b) AWI, where $\Omega_4= \omega _ 4- \omega _ {ml} $, $ L _ 4 = \alpha _ {40} ^ {-1}
$, $ \alpha _ {40} = \alpha _ {4}(G_i=0,\Omega_4=0)$, $G_{10}=100$, $G_{30}=40$ MHz,
$\Omega_1= \omega _ 1- \omega _ {gl}=0$ and $\Omega_3= \omega _ 3- \omega _ {mn}=100$
MHz.}
\end{figure}
If $ E _ {1,3} $ are homogeneous along $z$ (e.g., at the expense of saturation), the
system of Eqs. (\ref {e12}) and (\ref {e13}) reduces to two coupled equations for $ E
_ {4} $ and $ E _ {2} $, where the medium parameters are homogeneous as well. Their
solution is:
\begin{eqnarray}
&E_2^*=\exp\left(-\dfrac{\alpha_2}{2}\,z-\beta z\right)\;
\bigg \{-i\dfrac{\gamma_2^*}{R} E_{40}\sinh(Rz)&\nonumber\\
&+E_{20}^*\left[ \cosh(Rz)+\dfrac{\beta}{R} \sinh(Rz)\right]\bigg \},&\nonumber\\
&E_4=\exp\left(-\dfrac{\alpha_4}{2}\,z +\beta z\right)\;\bigg \{i\dfrac{\gamma_4}{R}
E_{20}^*\sinh(Rz)&\nonumber\\
&+E_{40}\left[\cosh(Rz)-\dfrac{\beta}{R} \sinh(Rz)\right]\bigg \}.&\label{smopa}
\end{eqnarray}
Here $R=\sqrt{\beta^2+\gamma^2},\ \beta = \left[(\alpha_4-\alpha_2)/2+i\Delta
k\right]/2,\ \Delta k=\delta k_1+\delta k_3-\delta k_2-\delta k_4, \
\gamma^2=\gamma_2^*\gamma_4$,\ $\gamma_{4, 2}=\chi_{4, 2}E_1E_3$, and $E_{20}^*$ and
$E_{40}$ are input values (at $z=0$). The first terms in braces indicate FWM and
second -- optical parametic  amplification (OPA) processes. If either of the driving
fields is switched off, $ \gamma _ {4, 2} = 0 $, and the weak radiations are
described as $|I_{4, 2}|^2=|I_{40,20}|^2exp\{-\alpha_{4, 2}z\}$ ($I_{4,2}=
|E_{4,2}|^2$). Owing to FWM coupling, probe radiation $ E _ 4 $ generates a wave
$E_2$ close to  $ \omega _ {gn} $, which in turn contributes to $E_4$ because of FWM.
This process results in correlated propagation of two waves along the medium. A gain
or absorption of any of them influences the propagation features of the other. If an
absorption (gain) exceeds the rate of the FWM conversion $(|\gamma^2|/\beta^2\ll 1)$,
$\Delta k=0$; if $E_{20}=0$, and $E_{40}\neq 0$, we obtain at $z=L$ the result that
$I_4/I_{40}=|\exp(-\alpha_4L/2)+ [{\gamma^2}/{(2\beta)^2}]\left[\exp(g_2L/2)-
\exp(-\alpha_4L/2)\right]|^2$, where $g_2\equiv -\alpha_2 $. Alternatively, if
$E_{40}=0$, and $E_{20}\neq 0$, $\eta_4 =I_4/I_{20}=[{|\gamma_4|^2}/{(2\beta)^2}]
\left|\exp(g_2L/2)- \exp(-\alpha_4 L/2)\right|^2$. One can see that achieving gain
requires large optical lengths $L$ and significant Stokes gain on the transition $gn$
([$ \exp (g _ 2L/2)] \gg | {(2\beta) ^ 2} / {\gamma ^ 2} | $), as well as effective
FWM  at both $\omega_2$ and $\omega_4$. The dependence $ I _ 4(L)$ is predetermined
by the sign of $ Im \gamma _ {4,2} $ and  $ Re \gamma _ {4,2} $.


The important feature of the far-from-degenerate interaction is that the magnitude
and the sign of the multiphoton resonance detunings and, consequently, of the
amplitude and phase of the lower-state coherence $\rho_{nl}$ differ for molecules at
different velocities because of the Doppler shifts. Such is not the case in
near-degenerate schemes. The interference of elementary quantum pathways, with
Maxwell's velocity distribution and saturation effects taken into account, results in
a nontrivial dependence of the macroscopic parameters on the intensities of the
driving fields and on the frequency detunings from the resonances \cite{Mys}. A
density matrix solution was found exactly with respect to $E_{1,3}$ and to a first
approximation for $E_{4,2}$. We use these formulas here for numerical averaging over
velocities, for analysis of the quantities $ \alpha _ {4,2}$ and $\gamma _ {4,2}$,
and also to obtain a numerical solution of the system of Eqs. (\ref {e12})-(\ref
{e13}), with the inhomogeneity of the coefficients taken into account. We stress that
the effect under consideration is AWI rather than conventional OPA accompanied by
absorption and Stokes gain, because the quantum interference involved in resonant
schemes is so crucial role that thinking in terms of the Manley-Raw conservation law
would be misleading \cite{TAR}.

The main outcomes of the simulations the conditions of the experiment are illustrated
in Figs. 1(b), 2 and 3. ~ The transitions  of Fig. 1(a) and relaxation parameters are
attributed to those of $Na_2$: $\lambda_{1-4}$ = 655, 756, 532, 480 nm,  $ \Gamma _
{m,\ g,\ n}$ = 260, 200, 30,  $ \gamma _ {mn,\ ml,\ gn,\ gl}$ = 24, 20, 10, 40, $
\Gamma _ {nm,\ lm}$ = 110, $ \Gamma _ {gm}$ = 130, and $ \Gamma _ {ng,\ lg}$ = 140
(all in $ 10 ^ {6} \times s ^ {-1} $). Here $ \Gamma _ {i}$ is the population, $
\Gamma _ {ij}$ are the coherence, and $ \gamma _ {ij}$ are the spontaneous interlevel
relaxation rates.  At $T=450$ $^\circ$C, the Doppler FWHM of the transition at $
\lambda _ 4$ is  1.7 GHz, and the Boltzmann population of level $ n $ is 2\% of that
of level $ l $. The Rabi frequencies $G_1=E_1d_{lg}/2\hbar$ and
$G_3=E_3d_{nm}/2\hbar$ of $\sim 100$ MHz correspond to 100-mW beams focused on a spot
with sizes of a few parts of a millimeter, i.e. one photon per several molecules.
\begin{figure}[htb]
\epsfysize=5.5cm \center{\leavevmode\epsfbox{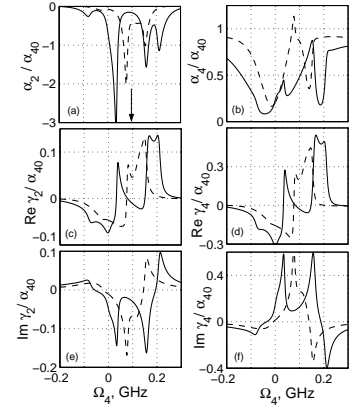}}~\caption{(a)~Macroscopic~Stokes~
gain,~(b)~ absorption,~ and~(c-f)~cross-coupling~indices; $\Omega_2\equiv
\omega_2-\omega_{gn} = \Omega_1+ \Omega_3 - \Omega_{4}$. Solid curves, same
parameters as in Fig. 1(b); dashed curves, $G_1=43$ MHz, $G_3=39$ MHz (corresponding
to $L/L_4=20$, where the gain reaches its maximum value).}
\end{figure}
\begin{figure}[htb]
\epsfysize=2.8cm \center{\leavevmode\epsfbox{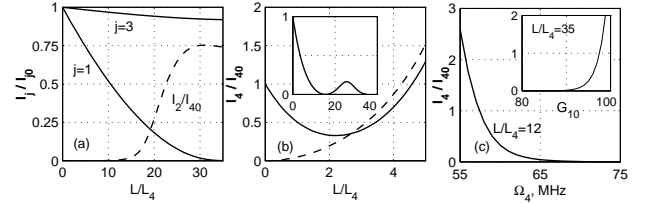}} \caption{(a,b) Coupled fields
versus optical length ($I_2$ is reduced by $10^4$), (c) CQC optical switching. For
(a) and (b) and for the inset in (c), $\Omega_4=160$ MHz, and for the inset in (b),
$\Omega_1=\Omega_2=\Omega_3=\Omega_4=0$, where the remaining parameters are the same
as in Fig. 1(b). [(c) $I_4/I_{40}= 5\times 10^{-4}$ at $\Omega _ 4 = 75 $ MHz and
$I_4/I_{40}= 10^{-2}$ at $\Omega _ 4 = 69 $ MHz; inset in (c), $I_4/I_{40}= 10^{-4}$
at $G_{10} = 80$ MHz and $I_4/I_{40}= 10^{-2}$ at $G_{10} =89$ MHz.]}
\end{figure}
\noindent However, the presence of such fields give rise to substantial modification
of both absorption and gain [Figs. 2(a) and 2(b)] and the cross-coupling [Figs.
2(c)-2(f)] parameters. Significant amplification in one of the nonlinear resonances
in $\alpha_2$ [Fig. 2 (a)] is accompanied by absorption peaks and transparency windows
in $\alpha_4$ [Fig. 2(b)]. The resonances in $\alpha_2$ do not coincide with a Raman
resonance [labeled with an arrow in Fig. 2(a)]. The modules of the dressed
velocity-averaged parameters $\gamma_{4,2}$ differ by approximately factor of four,
whereas their imaginary parts take on even different signs [as in Fig. 2(c)-2(f)].
This behavior is in marked contrast to that of solid-state and off-resonant nonlinear
optics.

The inhomogeneity of the driving fields [Fig. 3(a)] gives rise to a significant
change of the material parameters along the medium (dashed curves in Fig. 2), so
$\alpha_4$ may even increase above its value in the weak-field limit. The interplay
of these effects determines the spatial dynamics, optimum parameters, and achievable
gain [Fig. 1(b)]. Along a substantial medium length, the probe field is only being
depleted [Fig. 3(b)].  Its growth begins at the length where the generated and
enhanced field $E_2$ [dashed curves in Fig. 3(b)] becomes comparable with $E_{40}$.
The simulations explicitly reveal that the fully resonant conditions explored in Ref.
8 are far from optimal [inset in Fig. 3(b)], and most probably the gain reported in
Ref. 8 is a misinterpretation of the experiment. The maximum gain in Fig. 1(b) is
1050, which is well above the characteristic threshold for self-oscillation to be
established inside the optical cavity from the spontaneous radiation. This gain can
readily be increased further to the mirrorless oscillation level. Both linear and
laser-induced nonlinear dispersion inhomogeneous along the medium are taken into
account in Figs. 1(b) and 3.  Our results also demonstrate that the problem of AWI in
similar schemes may not be reduced to the condition of a sign change of $\alpha_4$,
as was done in the research reported in Ref. 11.  The solid curve in Fig. 3(b) shows
that there is an optical thickness controlled by the driving radiations whose small
variation results in a switching from the absorption regime to transparency and
further to amplification. Figure 3(c) presents the possibility of controlling this
switching with a small change of either the frequency of the probe radiation or the
intensity of the driving radiation (inset, Fig. 3).
Obviously, the same processes can be employed for
generating and manipulating large dispersion without the accompanying depletion of
radiation.
The required intensity can be further decreased in identical but  more favorable
atomic schemes. Owing to the generated molecular coherence, fields $E_4$ and $E_2$
may possess nearly perfect quantum correlations that yield almost complete
squeezing\cite{QN}.

We thank S.A. Myslivets for assistance in the numerical simulations and the U.S.
National Research Council - National Academy of Sciences for support through the
international Collaboration in Basic Science and Engineering program. AKP
acknowledges stimulating discussions with B. Wellegehausen and B. Chichkov, and
support from the International Association of the European Community for the
promotion of co-operation with scientists from the New Independent States of the
former Soviet Union  (grant INTAS-99-19), the Russian and Krasnoyarsk Regional
Foundations for Fundamental Research (grant 99-02-39003) and from the Center on
Fundamental Natural Sciences at St. Petersburg University (Russia) (grant 97-5.2-61).
Thomas F. George's e-mail address is
\href{mailto:tgeorge@uwsp.edu}{tgeorge@uwsp.edu}; that of Alexander K. Popov is
\href{mailto:popov@ksc.krasn.ru}{popov@ksc.krasn.ru}.

\end{document}